\documentclass[journal]{IEEEtran}

\usepackage{amsfonts,amsmath,amssymb}
\usepackage{dsfont}
\usepackage{cite}
\usepackage{graphicx}
\usepackage{url}
\usepackage{caption}
\usepackage{subcaption}
\usepackage{bm}
\usepackage{bbding}
\usepackage{amsmath}
\usepackage{enumerate}
\usepackage{multirow}
\usepackage{algorithm}
\usepackage{algorithmic}
\usepackage{array}
\usepackage{CJK}
\usepackage{amsthm}

\allowdisplaybreaks

\begin{document}
\title{Coverage Probability Analysis of UAV Cellular Networks in Urban Environments}
\author{\IEEEauthorblockN{Lai Zhou$^1$, Zhi Yang$^2$, Shidong Zhou$^2$, Wei Zhang$^3$}

\IEEEauthorblockA{
{$^1$}
Department of Engineering Physics, Tsinghua University, Beijing, China\\
{$^2$}
Department of Electronic Engineering, Tsinghua University, Beijing, China\\
{$^3$}
The University of New South Wales, Sydney, Australia\\
Email: $\{$zhoulai13,yang-z15$\}$@mails.tsinghua.edu.cn, zhousd@mail.tsinghua.edu.cn, w.zhang@unsw.edu.au}}

% make the title area
\IEEEoverridecommandlockouts

\maketitle
\begin{abstract}
In this paper, we study coverage probabilities of the UAV-assisted cellular network modeled by 2-dimension (2D) Poisson point process. The cellular user is assumed to connect to the nearest aerial base station. We derive the explicit expressions for the downlink coverage probability for the Rayleigh fading channel. Furthermore, we explore the behavior of performance when taking the property of air-to-ground channel into consideration. Our analytical and numerical results show that the coverage probability is affected by UAV height, pathloss exponent and UAV density. To maximize the coverage probability, the optimal height and density of UAVs are studied, which could be beneficial for the UAV deployment design.
\\
\end{abstract}

%\keywords{Stochastic geometry, UAV, coverage probability, air-to-ground channel, cellular network.}


%\begin{keywords}
%Stochastic geometry, UAV, coverage probability, air-to-ground channel, cellular network.
%\end{keywords}

\section{Introduction}

Enabled by the improvement in control, communication and miniaturization of the hardware, unmanned aerial vehicles (UAV) have been attracting significant attentions in the recent years, which motivate some commercial applications, e.g., transportation of goods, disaster relief and traffic monitoring, etc. \cite{magazine1}\cite{life1} By taking the advantage of flexible deployment and good connectivity condition in a certain height, the UAV-assisted wireless communication could be used to support network in some scenarios, e.g., temporary deployment to provide regional coverage in the wake of disasters, capacity enhancement in the occasional demand of super dense base stations, and aerial relay for device-to-device (D2D) communications on the ground \cite{cellular1} $-$\cite{relay1}.  Hence, the necessity and benefit motivate us to study the performance of the UAV-assisted cellular network, especially for its property in some typical propagation scenarios.

\subsection{Related work and Motivation}

Considering the importance of channel model in the system design, much research effort has been devoted to the air-to-ground (AG) channels at high-altitude \cite{mea1}, and recent applications to the low-altitude \cite{mea2}\cite{mea3}\cite{mea4}. Different from the widely used channel model based on the simulation result \cite{sim1}, the measurement in \cite{mea3} shows that the pathloss exponent (PLE) decreases with an increase in the height. In addition, \cite{mea4} shows that it is also related to other factors from the surrounding buildings and trees. These channel properties would have impact on the performance of the UAV-assisted cellular networks. Another recent work shows that the probability of line-of-sight (LOS) for AG channel is sensitive to the height and elevation angle \cite{LOS}. Based on the channel properties, \cite{cellular1}\cite{cellular2}\cite{cellular3} study the performance of the drone small cells (DSC), but most of the results are obtained from simulations.

In order to model the deployments of base stations or mobile users with tractability, stochastic geometry theory has been widely used to analyze \textit{ad hoc} and cellular network \cite{ppp1} $-$\cite{ppp3}. By modeling the UAV as homogeneous Poisson Point Process (PPP), \cite{JSAC} extends the wireless networks to 3-dimension (3D) space with finite height and derives the optimal density of DSCs. Recent work \cite{TWC}, studies the average downlink coverage in a finite 3D network with homogeneous binomial Poisson process (BPP) and concludes some potential properties, e.g., the coverage probability is reduced with a decrease in the PLE. However, The work does not consider the effects of noise, which may result in different conclusion. Moreover, none of those studies consider the impact of AG channel.

\subsection{Contributions}
Considering the demand of dense base station deployment in the urban scenario, we focus on the performance of UAV-assisted cellular networks based on the typical channel property. The main contributions of this paper are listed below.

\emph{1) Coverage probability:} The Nakagami-m fading could be used to express the general small scale fading \cite{ppp2}, and the special cases are Rician fading in the LOS scenario and Rayleigh fading in the Non-line-of-sight (NLOS) scenario. We derive the general expressions for the downlink coverage probability based on Nakagami-m fading channel, and then derive the explicit expressions for Rayleigh fading channel.

\emph{2) Performance analysis based on the AG channel:} With the explicit expressions for Rayleigh fading channel, we analyze the effects of channel characteristics and obtain some important performance trends in terms of the UAV height, PLE, and UAV density. The analysis result would be totally different in the case of low and high signal-to-noise ratio (SNR), respectively. In high SNR, the coverage probability degrades as the PLE decreases. But in low SNR, the coverage probability degrades with an increase in the PLE.

The rest of the paper is organized as follows, Section II introduces the system model. Section III derives the coverage probability for different types of small scale fading channel. The effects of the AG channel on the coverage probability are analyzed and discussed in Section V. Finally, Section VI concludes the paper.

\section{System model}

\begin{figure}
    \centering
    \includegraphics[width=3.3in]{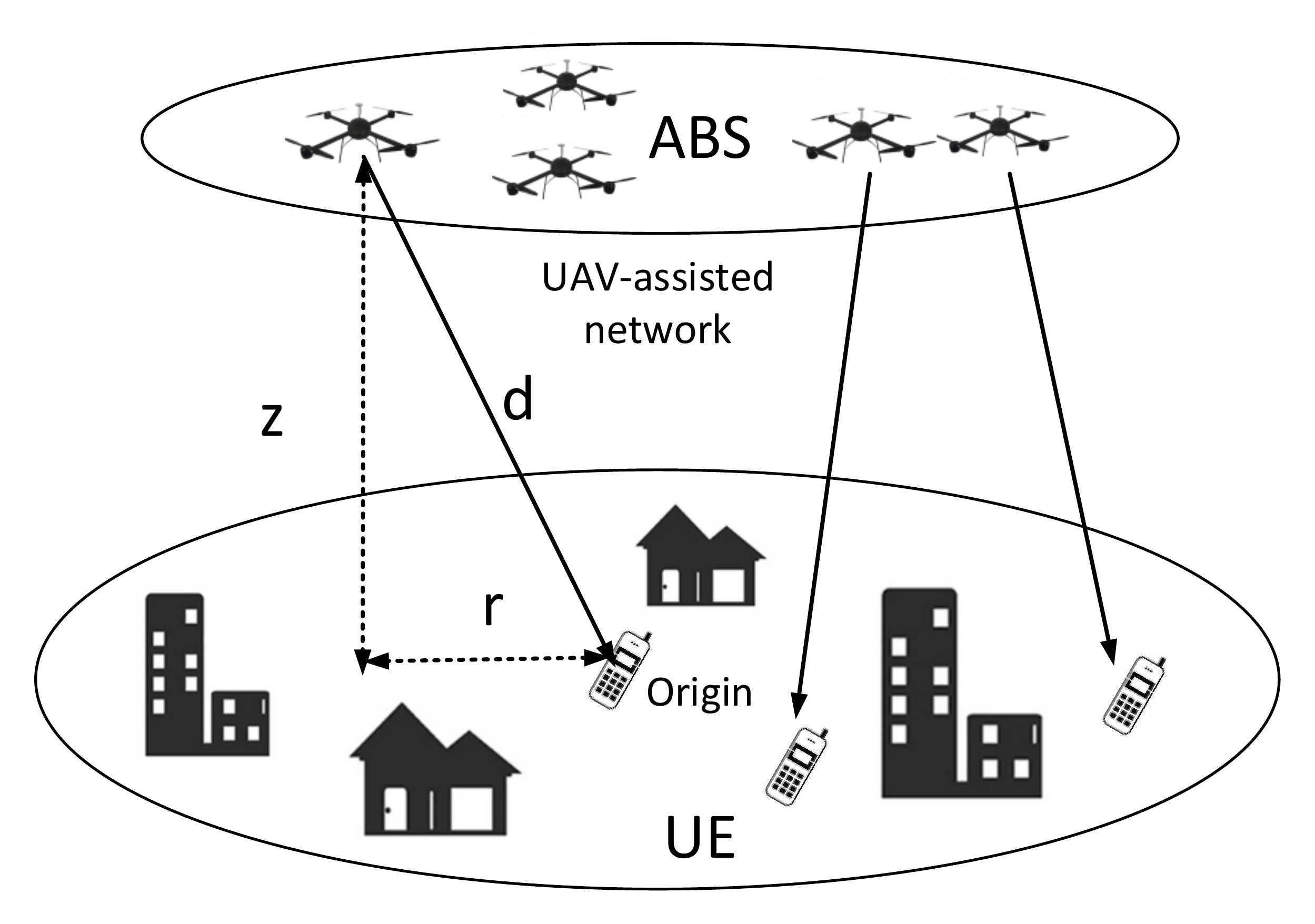}
    \vspace{-3mm}
    \caption{UAV-assisted cellular networks.}
    \label{fig1}
    \vspace{-3mm}
\end{figure}

As shown in Fig. \ref{fig1}, the UAVs are deployed to act as the aerial base stations (ABSs) in 2D space. The UAV follow a 2D-PPP ($X_i \in \Phi$) with density $\lambda$ in the infinite space $\mathbf{U}$, that is, $\mathbf{U}=\{(x,y,z):x,y \in \mathbb{R}, z = L_0 \}$. For simplicity, we assume that all the UAVs are deployed at the same height $L_0$, and $L_{min}\leq L_0 \leq L_{max}$. Considering the regulation about the safety altitude for commercial UAV, $L_{min}$ and $L_{max}$ are the minimum and maximum of the altitude, respectively. The ABSs could be used to support terrestrial networks in overloaded situations, such as for the purpose of disaster management and temporary coverage enhancement in some hotspots, which we define as the UAV-assisted cellular networks. However, the user equipment (UE) on the ground would suffer from the interference from other ABSs, which may have impact on the performance of the networks.

The channel between the UAV and UE is the AG channel, which suffers from the path loss and the small scale fading. In this work, we use the Stanford University Interim (SUI) Model \cite{mea6}. The channel gain is expressed as
\begin{equation} \label{eq11}
Gain=h\cdot  K_0  (d/d_0)^{-n(z)},
\vspace{0mm}
\end{equation}
where
\begin{equation} \label{eq1}
n(z) = \emph{max}(a-b z+c/z,2),
\vspace{0mm}
\end{equation}
$d$ is the propagation distance, $K_0$ is the received power at a reference distance $d_0$, $n(z)$ is the PLE at a height of $z$, and $h$ is the small scale fading in a certain distribution, e.g., Nakagami distribution and Rayleigh distribution. Parameters a, b, and c are constants to model the terrain types, and the suitable parameters are $a = 4.6,b=0.0075,c=12.6$ \cite{mea6}. The empirically based model combines the LOS and NLOS scenario together, and the PLE is about 2 at a high altitude since there are few blockages.

Due to Slivnyak's theorem \cite{ppp1}, the UE at the origin is used to analyze the performance of the cellular network. We assume that the UE would communication with the closest ABS, so the probability density function (PDF) of $r$ (the projection of the propagation distance on the ground) is \cite{ppp2},
\begin{equation} \label{eq2}
f(r) = 2  \lambda \pi r e^{- \lambda \pi r^2}.
\vspace{0mm}
\end{equation}
For simplicity, all ABSs are assumed to transmit at the same power level $P_t$, and the received signal-to-interference-plus-noise ratio (SINR) can be expressed as,
\begin{equation} \label{eq3}
\begin{aligned}
SINR
& = \frac{P_t\cdot h K_0(d/d_0)^{-n(z)}}{N+\sum_{d_i \in \Phi \backslash \{0\} }{P_t\cdot h_i K_0(d_i/d_0)^{-n(z)} }} \\
& = \frac{ h (d/d_0)^{-n(z)}}{\beta_0 +\sum_{d_i \in \Phi \backslash \{0\} }{ h_i (d_i/d_0)^{-n(z)} }}
\vspace{0mm}
\end{aligned}
\end{equation}
where $d = \sqrt{r^2+z^2}$, the noise power is assumed to be constant with value $N$, and $I=\sum_{d_i \in \Phi \backslash \{0\} }{h_i (d_i/d_0)^{-n(z)} }$ is the sum of the normalized interference. It should be noted that, $\frac{1}{\beta_0} = \frac{P_t K_0}{N}$ is defined to be the received SNR at a reference distance $d_0$. The UE connects with the ABS successfully only when the SINR is larger than a certain threshold $\theta$, and the coverage probability averaged over the plane is,
\begin{equation} \label{eq4}
\begin{aligned}
\textbf{P}(\theta,z)
& = \mathbb{E}_r[\mathbb{P}(SINR>\theta)\mid r] \\
& = \int_0^\infty \mathbb{P}(SINR>\theta \mid r)f(r)dr.
\end{aligned}
\vspace{0mm}
\end{equation}

\section{Coverage probability}

The Rician distribution is usually used to model the LOS scenario with dominant paths, and the non-dominant multipaths which are severely affected by fading are modeled by Rayleigh distribution. Since Nakagami-m distribution is a universal model suitable for various conditions, we use the general fading distribution to derive the general expression of coverage probability. Note that the gain $h$ follows a Gamma distribution with unit mean as \cite{molisch}
\begin{equation} \label{eq5}
f_G(h) = \frac{m^mh^{m-1}}{\Gamma(m)}e^{-mh},
\vspace{0mm}
\end{equation}
which we denote as $h \sim G(m,1/m)$, the $m$ factor can be computed from $K$ factor in the Rician fading ($m>1$) as \cite{molisch}
\begin{equation} \label{eq6}
m=\frac{K^2+2K+1}{2K+1},
\vspace{0mm}
\end{equation}
and the Rayleigh fading occurs when $m=1$ and the power gain follows the exponential distribution as $h \sim \textrm{exp}(1)$. The general expression for coverage probability is
\begin{equation} \label{eq7}
\begin{aligned}
\textbf{P}(\theta,z)
& = \int_0^\infty \mathbb{P}(\frac{h(d/d_0)^{-n(z)}}{\beta_0+I} >\theta \mid r) \cdot 2 \lambda \pi r e^{-\lambda \pi  r^2} dr \\
& \overset{(a)}{=} \lambda \pi \int_0^\infty \mathbb{E}_I \left[ \frac{\Gamma(m,m \mu)}{\Gamma(m)} \mid r \right] e^{-\lambda \pi  v} dv,
\vspace{0mm}
\end{aligned}
\end{equation}
where $\mu = \theta(\beta_0+I)(d/d_0)^{n(z)}$, $d = \sqrt{v+z^2}$, (a) follows because of the Gamma distribution as (\ref{eq5}) and the substitution $r^2 \rightarrow v$, $\Gamma(s,x)=\int_x^\infty t^{s-1}e^{-t}dt$ denotes the incomplete Gamma function, and $\Gamma(s) = \int_0^\infty t^{s-1}e^{-t}dt$ denotes the standard Gamma function. Since the complexity of the expression forbids any further analysis, the accurate expression for the Rayleigh fading channel ($m$ = 1) is derived as
\begin{equation} \label{eq8}
\textbf{P}(\theta,z) = \lambda \pi e^{-\lambda \pi \rho z^2} \int_0^\infty e^{-\lambda \pi (1+\rho)v-\theta \beta_0 (\frac{v+z^2}{d_0^2})^{n(z)/2}}dv
\vspace{0mm}
\end{equation}
where
\begin{equation} \label{eq81}
\rho = \theta^{2/n(z)} \int_{\theta^{-2/n(z)}}^\infty \frac{1}{1+x^{n(z)/2}}dx,
\vspace{0mm}
\end{equation}
and the theoretical expression could be further simplified to
\begin{equation} \label{eq11}
\textbf{P}(\theta,z) = \frac{\textrm{exp}(-\lambda \pi \rho z^2)}{1+\rho},
\vspace{0mm}
\end{equation}
with the assumption of no noise.

The proof of Eq. (\ref{eq8}) is given as follows. Conditioning on the nearest ABS at a distance $d$ from the typical UE (the corresponding projection distance is $r$), the coverage probability is
\begin{equation} \label{eq9}
\begin{aligned}
& \mathbb{P}(\frac{h(\frac{d}{d_0})^{-n(z)}}{\beta_0+I} >\theta \mid r) \\
& = \mathbb{P}(h>\theta (\frac{d}{d_0})^{n(z)}(\beta_0+I) \mid r) \\
& \overset{(b)}{=} \mathbb{E}_I \left[ \textrm{exp}(-\theta (\frac{d}{d_0})^{n(z)}\beta_0-\theta (\frac{d}{d_0})^{n(z)}I) \mid r\right] \\
& =  \textrm{exp}\left[-\theta \beta_0 (\frac{d}{d_0})^{n(z)} \right] \mathcal{L}_I \left[ \theta (\frac{d}{d_0})^{n(z)} \right]
\vspace{0mm}
\end{aligned}
\end{equation}
where (b) follows because $h$ is exponential distributed with unit mean, and $\mathcal{L}_I \left[ \theta (d/d_0)^{n(z)} \right]$ is the Laplace transform of the interference which can be further derived as,
\begin{equation} \label{eq10}
\begin{aligned}
\mathcal{L}_I
& = \mathbb{E}_I \left[ \textrm{exp}(-\theta (d/d_0)^{n(z)}I) \mid r\right] \\
& = \mathbb{E}_{\Phi,h_i} \left[   \prod_{d_i\in \Phi \backslash \{0\}} \textrm{exp}(-\theta d^{n(z)}h_i d_i^{-n(z)})   \right] \\
& \overset{(c)}{=} \mathbb{E}_{\Phi} \left[   \prod_{d_i\in \Phi \backslash \{0\}} \frac{1}{1+\theta d^{n(z)}d_i^{-n(z)}} \right] \\
& \overset{(d)}{=} \textrm{exp}\left[-2 \lambda \pi \int_r^\infty (1-\frac{1}{1+\theta d^{n(z)}(\sqrt{v^2+z^2})^{-n(z)}}) v dv\right] \\
& \overset{(e)}{=} \textrm{exp} \left[   -\lambda \pi d^2 \theta^{2/n(z)} \int_{\theta^{-2/n(z)}}^\infty \frac{1}{1+x^{n(z)/2}}dx  \right]
\vspace{0mm}
\end{aligned}
\end{equation}
where (c) follows from the i.i.d. distribution of $h_i$, its independence from the point process $\Phi$, and its exponential distribution property. (d) follows from the probability generating function (PGFL) of the PPP \cite{ppp1}, and the integration limits expresses that the interference should be farther than the distance $d = \sqrt{r^2+z^2}$. (e) follows by using the change of variables $\frac{v^2+z^2}{\theta^{2/n(z)}d^2} \rightarrow x $. Plugging (\ref{eq2})(\ref{eq9})(\ref{eq10}) into (\ref{eq4}) with $r^2 \rightarrow v$ gives the desired result, and the proof of (\ref{eq8}) is complete.

\begin{figure}
    \centering
    \includegraphics[width=3.7in]{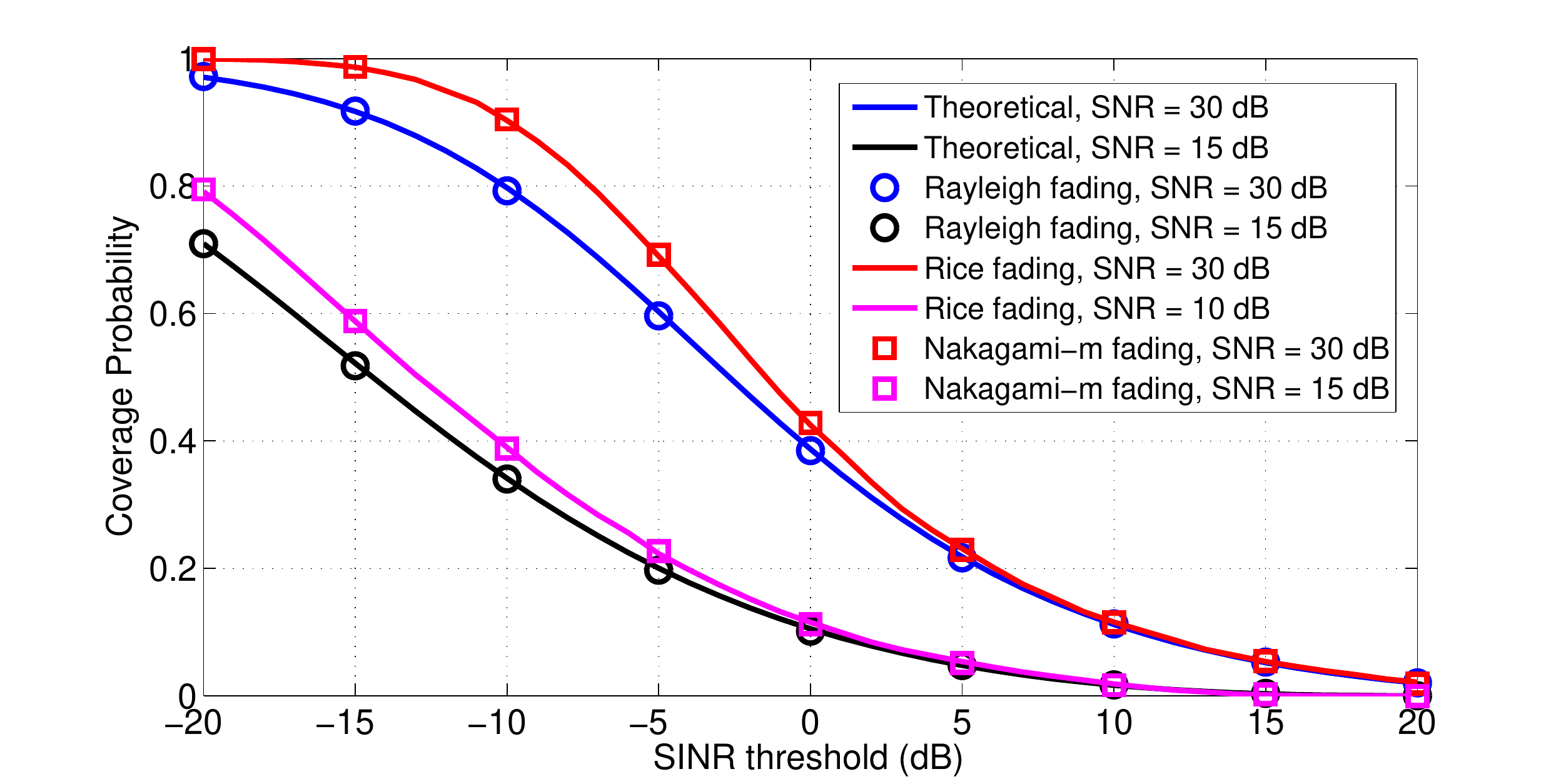}
    \caption{Comparison between different types of small scale fading, condition: PLE = 4, height = 100 m, $\lambda$ = 1$/\textrm{km}^2$, $d_0$ = 100 m.}
    \label{fig3}
    \vspace{-3mm}
\end{figure}

Fig. \ref{fig3} shows the coverage probability for different fading channels in a typical scenario. According to the SUI model, the PLE will be 4 at an altitude of 100 m. The $K$ factor in Rician fading is usually more than 10 dB in the suburb scenario \cite{mea1}\cite{mea2}, but less than 5 dB in the urban scenario because of blockage and multipath from the surroundings \cite{mea4}. In the simulation, the $K$ factor of Rician fading is set as 10 dB to evaluate the difference as much as possible, and the corresponding parameter of Nakagami-m fading is $m = 5.8$ as equation (\ref{eq6}). Note that the SNR is at a reference distance of $d_0$ = 100 m. We may conclude following notes:
\begin{enumerate}
\item With the Rayleigh fading channel, the theoretical result (\ref{eq8}) matches the simulation result very well;
\item Nakagami-m fading could be used to approximate the Rician fading;
\item The Rayleigh fading degrades the performance compared to the Rician fading, and the difference becomes less with lower SNR and higher SINR threshold.
\end{enumerate}

Hence, the expression for the Rayleigh fading channel could be used to approach the coverage probability, in particular for the urban scenario with small $K$ factor, e.g., 5 dB.

\section{performance Analysis and Discussion}

In this section, the average coverage probability (\ref{eq8}) is further analyzed with different system parameters, e.g., SNR, altitude, PLE, and UAV density.

\subsection{Altitude}
The PLE is assumed to be the typical value, i.e., 4, and the UAV density is $\lambda$ = 1$/\textrm{km}^2$. The expression of average coverage probability is derived from (\ref{eq8}) as,
\begin{equation} \label{eq12}
\begin{aligned}
\textbf{P}(\theta,z)
& = \lambda \pi e^{-\lambda \pi \rho z^2} \int_0^\infty e^{-\lambda \pi (1+\rho)v-\theta \beta_0 (\frac{v+z^2}{d_0^2})^2}dv \\
& = \frac{\lambda \pi^{\frac{3}{2}}d_0^2}{\sqrt{\theta \beta_0}} e^{\frac{\kappa^2}{2}+\lambda \pi z^2} Q(\kappa+\frac{z^2}{d_0^2}\sqrt{2\theta \beta_0}),
\vspace{0mm}
\end{aligned}
\end{equation}
where
\begin{equation} \label{eq121}
\rho = \theta^{1/2}(\frac{\pi}{2}-\textrm{arctan}(\theta^{-1/2})),
\vspace{0mm}
\end{equation}
\begin{equation} \label{eq122}
\kappa = \frac{\lambda \pi (1+\rho)d_0^2}{\sqrt{2\theta \beta_0}  },
\vspace{0mm}
\end{equation}
and $Q(\cdot)$ is the $Q$-function.

\begin{figure}\centering
    \includegraphics[width=3.7in]{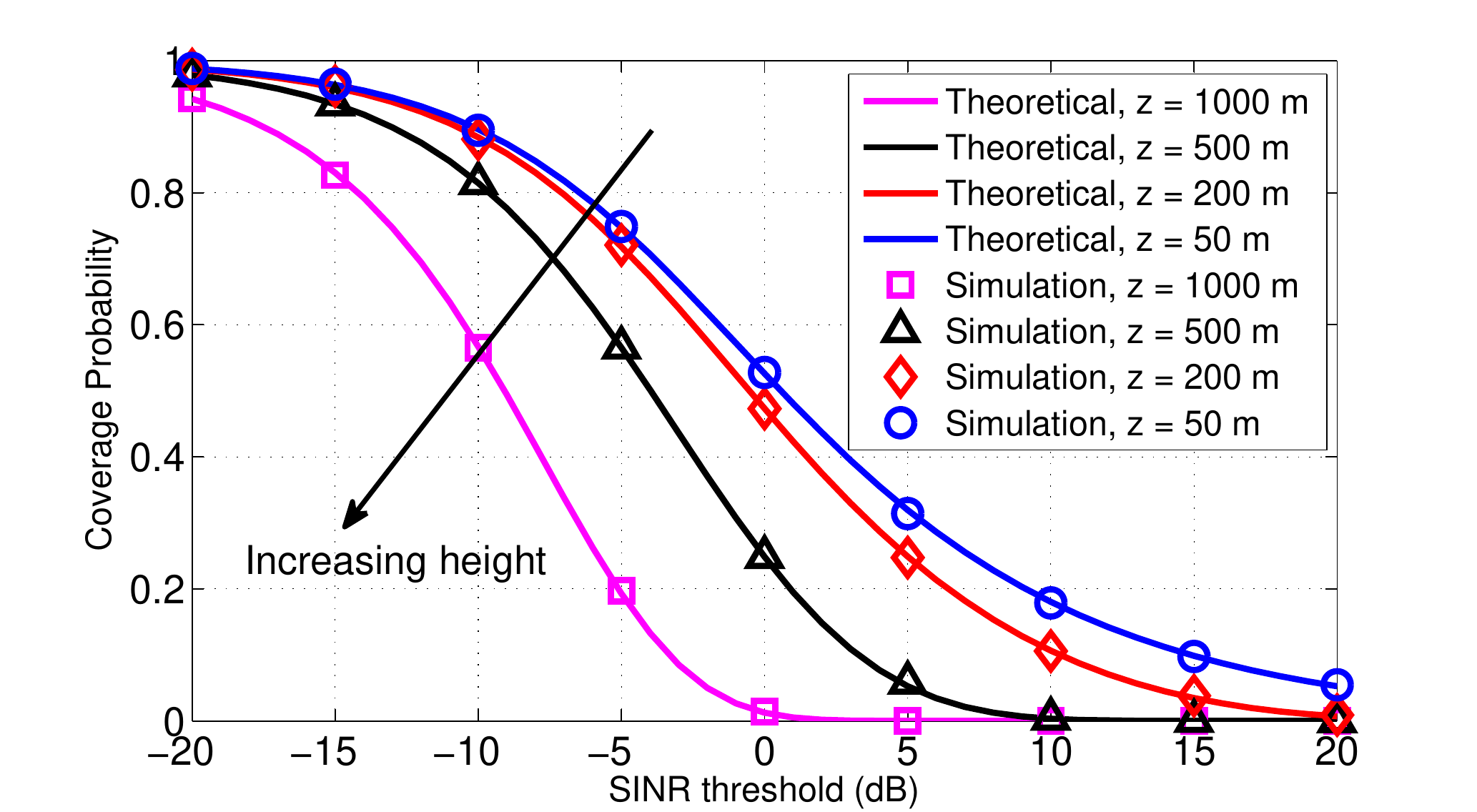}
    \caption{Comparison between different heights, condition: PLE = 4, SNR = 40 dB, $\lambda$ = 1$/\textrm{km}^2$, $d_0$ = 100 m.}
    \label{fig4}
    \vspace{-3mm}
\end{figure}

\begin{figure}\centering
    \includegraphics[width=3.7in]{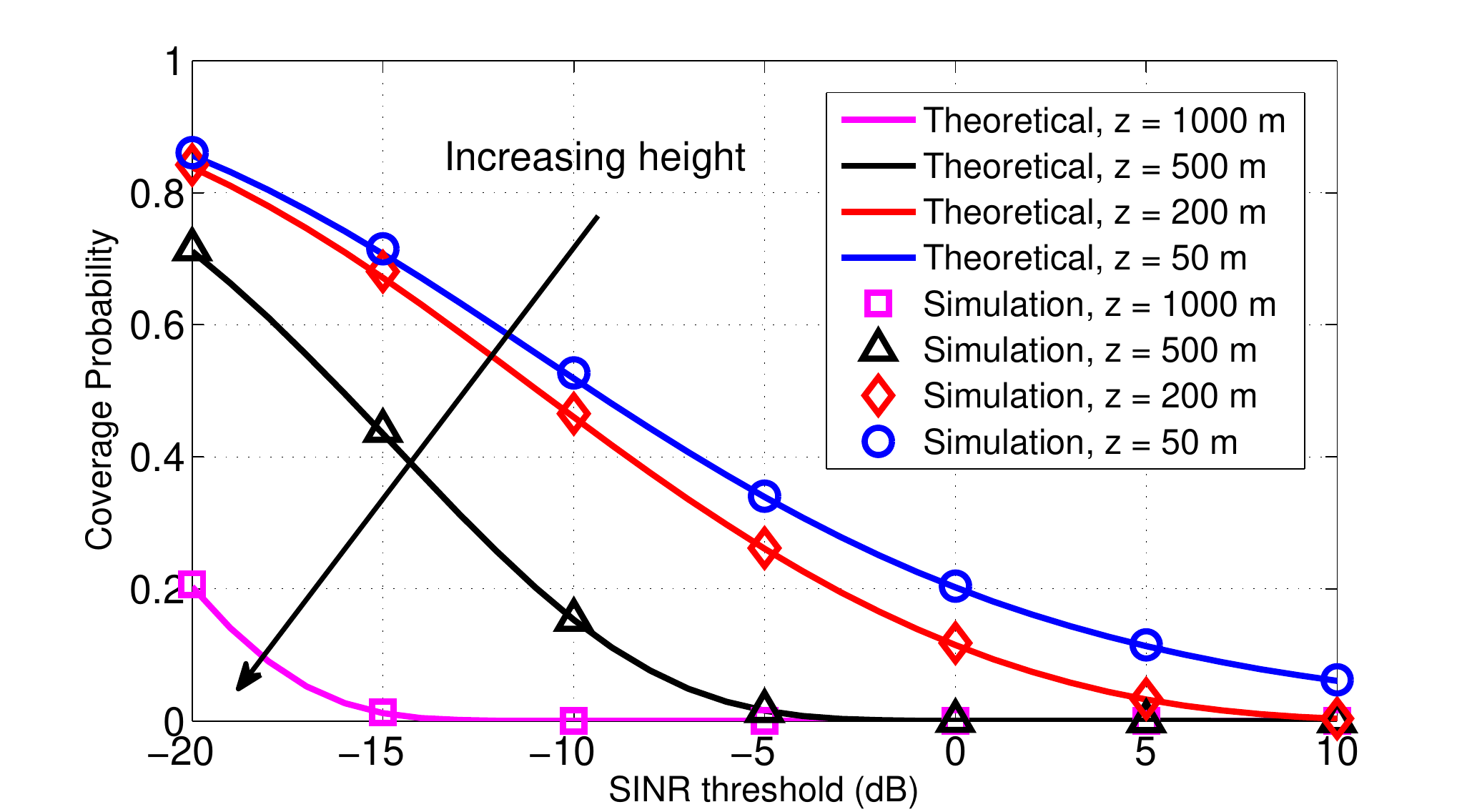}
    \caption{Comparison between different heights, condition: PLE = 4, SNR = 20 dB, $\lambda$ = 1$/\textrm{km}^2$, $d_0$ = 100 m.}
    \label{fig41}
    \vspace{-3mm}
\end{figure}

Fig. \ref{fig4} and Fig. \ref{fig41} show the theoretical and simulation results with the SNR of 40 dB and 20 dB, respectively.
It is observed that the coverage probability degrades with an increase in height in the two plots. In the high SNR condition, the noise is very small compared to the desired signal power, so the SINR mainly depends on the ratio between the signal power and interference. The relative separation between the serving and interfering ABS would degrade with an increase in the UAV height, so the SINR decreases hence the coverage. Meanwhile, the interference is negligible in the low SNR condition, and the increase of the height will lead to more pathloss, which worsens the signal power and hence the SINR and coverage.

\subsection{PLE}

Considering the limitation of the minimal height and the SUI model, e.g., $L_{min} = 20$ m, the range of PLE is from 2 to 5. Note that the height is assumed to be fixed here and the analysis of combining the PLE and height will be shown in Part C.

\begin{figure}
    \centering
    \includegraphics[width=3.7in]{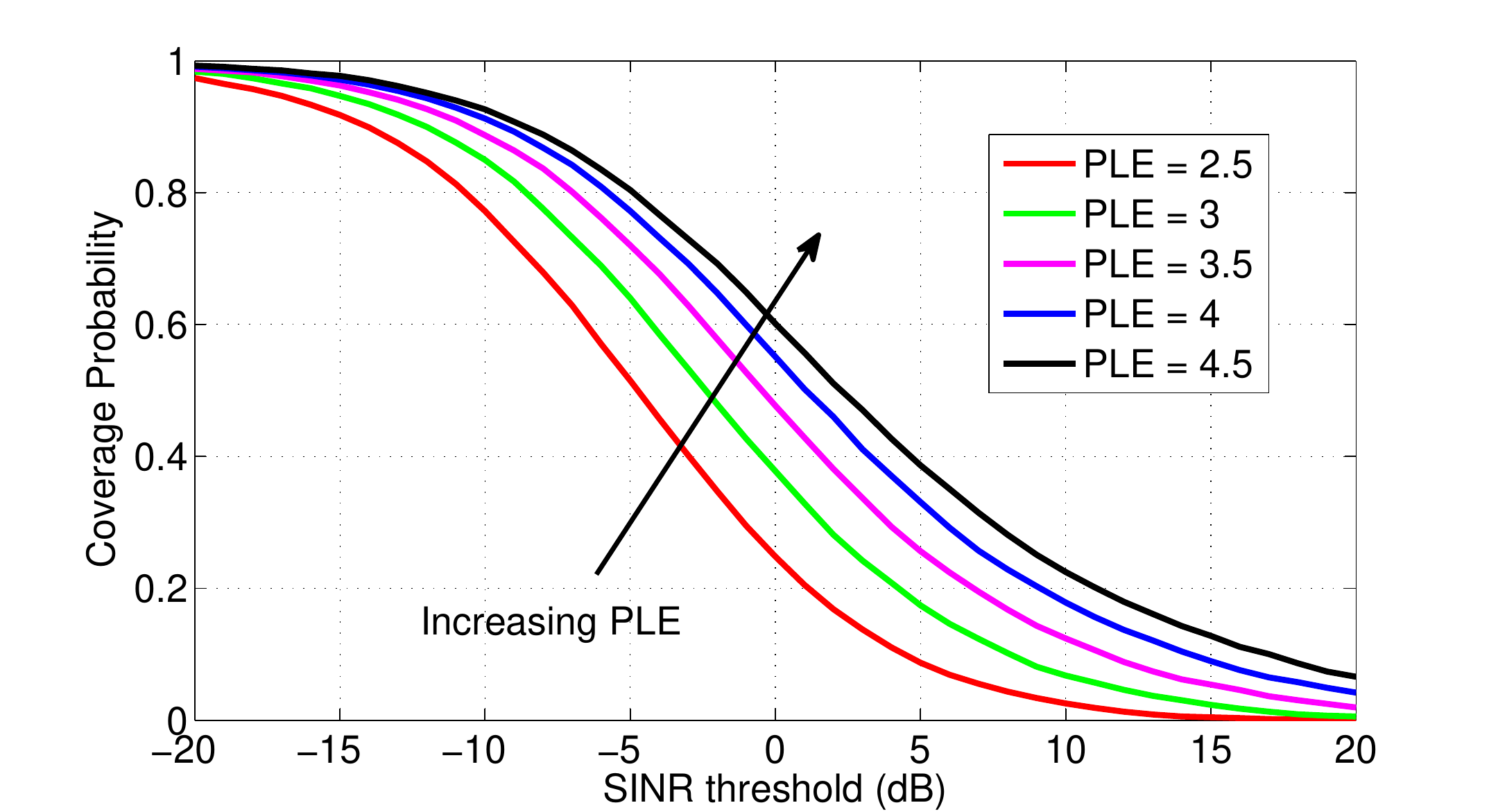}
    \caption{Comparison between different PLEs, condition: SNR = 50 dB, height = 100 m, $\lambda$ = 1$/\textrm{km}^2$, $d_0$ = 100 m.}
    \label{fig5}
    \vspace{-3mm}
\end{figure}

\begin{figure}
    \centering
    \includegraphics[width=3.7in]{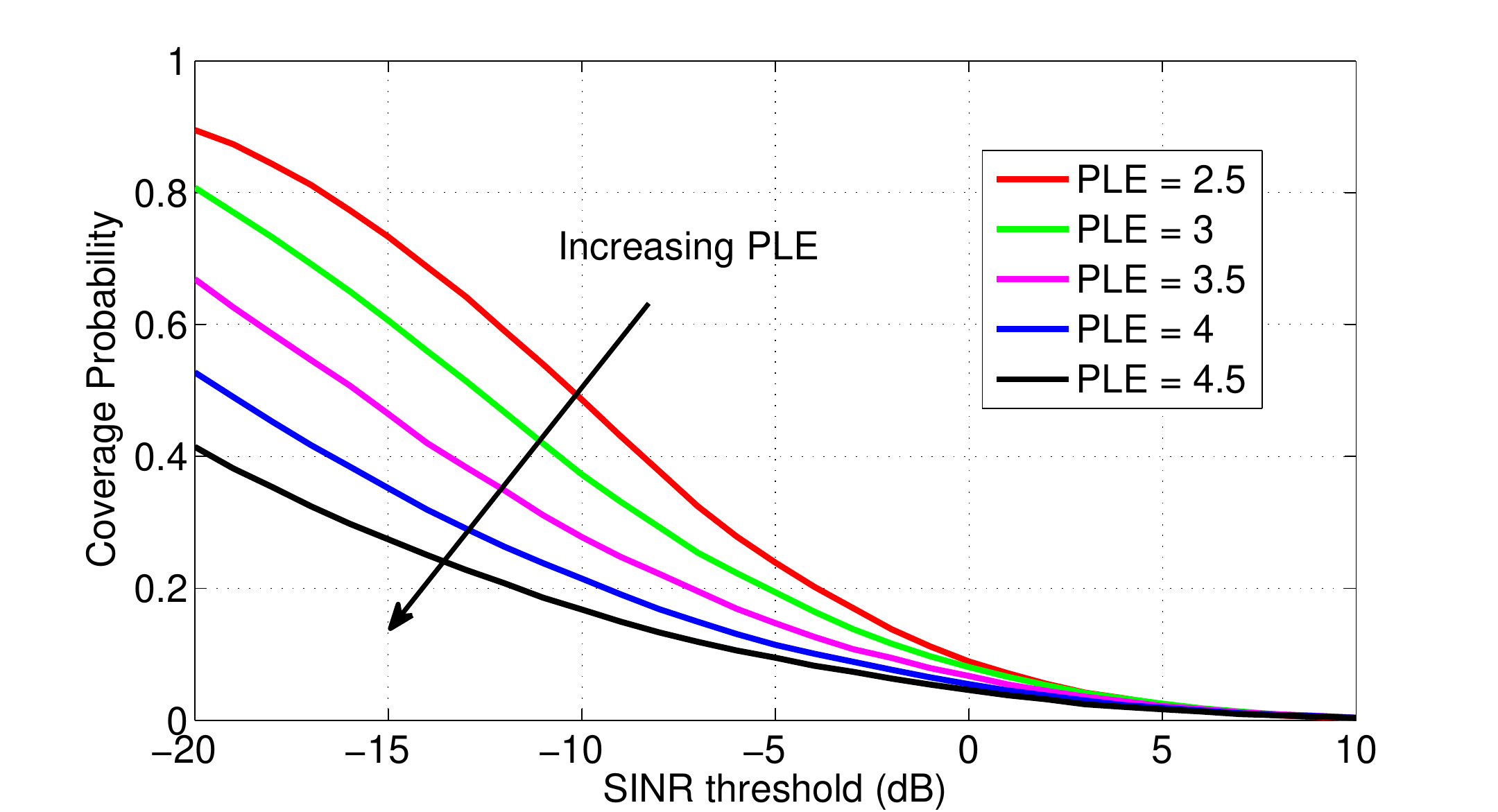}
    \caption{Comparison between different PLEs, condition: SNR = 10 dB, height = 100 m, $\lambda$ = 1$/\textrm{km}^2$, $d_0$ = 100 m.}
    \label{fig6}
    \vspace{-3mm}
\end{figure}

Fig. \ref{fig5} and Fig. \ref{fig6} show the simulation result with the high and low SNR, respectively. It can be observed that the increasing PLE has the opposite effect on the coverage probability. With the high SNR, both the signal and interference will suffer from more pathloss when the PLE increases, but the interference power decreases faster because of larger propagation distance, which improves the SINR and hence the coverage. On the contrary, in the low SNR condition the interference could be neglected $\rho = 0$, so the theoretical expression (\ref{eq8}) could be simplified to
\begin{equation} \label{eq123}
\textbf{P}(\theta,z) = \lambda \pi \int_0^\infty e^{-\lambda \pi v-\theta \beta_0 (\frac{v+z^2}{d_0^2})^{n/2}}dv.
\vspace{0mm}
\end{equation}
which implies that the coverage probability degrades as the PLE increase. Essentially, the increasing PLE decreases the received power of
the desired signal, thereby degrading the SINR and hence the coverage.

\subsection{Optimal Altitude}

In the AG channel, the PLE will decrease with an increase in the height of ABS, which improves the communication condition with less blockage. In the high SNR condition, increasing the height and reducing the PLE both lead to a decrease in the coverage probability. However, with the low SNR, the PLE and height will influence the coverage in the opposite direction, so there should be an optimal height to maximize the coverage probability.

\begin{figure}\centering
    \includegraphics[width=3.7in]{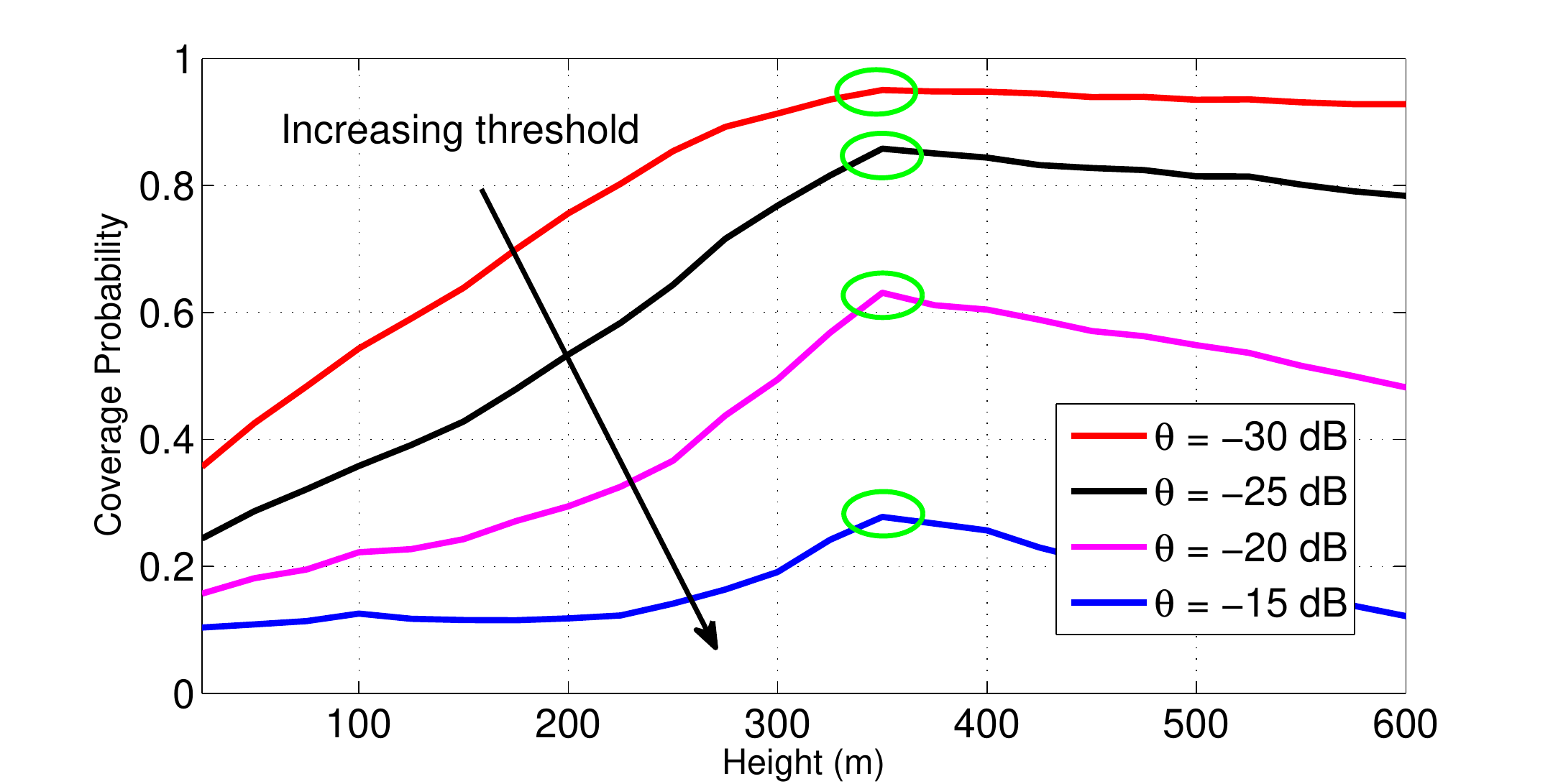}
    \caption{Simulation based on  SUI model in different SINR thresholds, condition: SNR = 0 dB, $\lambda$ = 1$/\textrm{km}^2$, $d_0$ = 100 m.}
    \label{fig7}
    \vspace{-3mm}
\end{figure}

\begin{figure}\centering
    \includegraphics[width=3.7in]{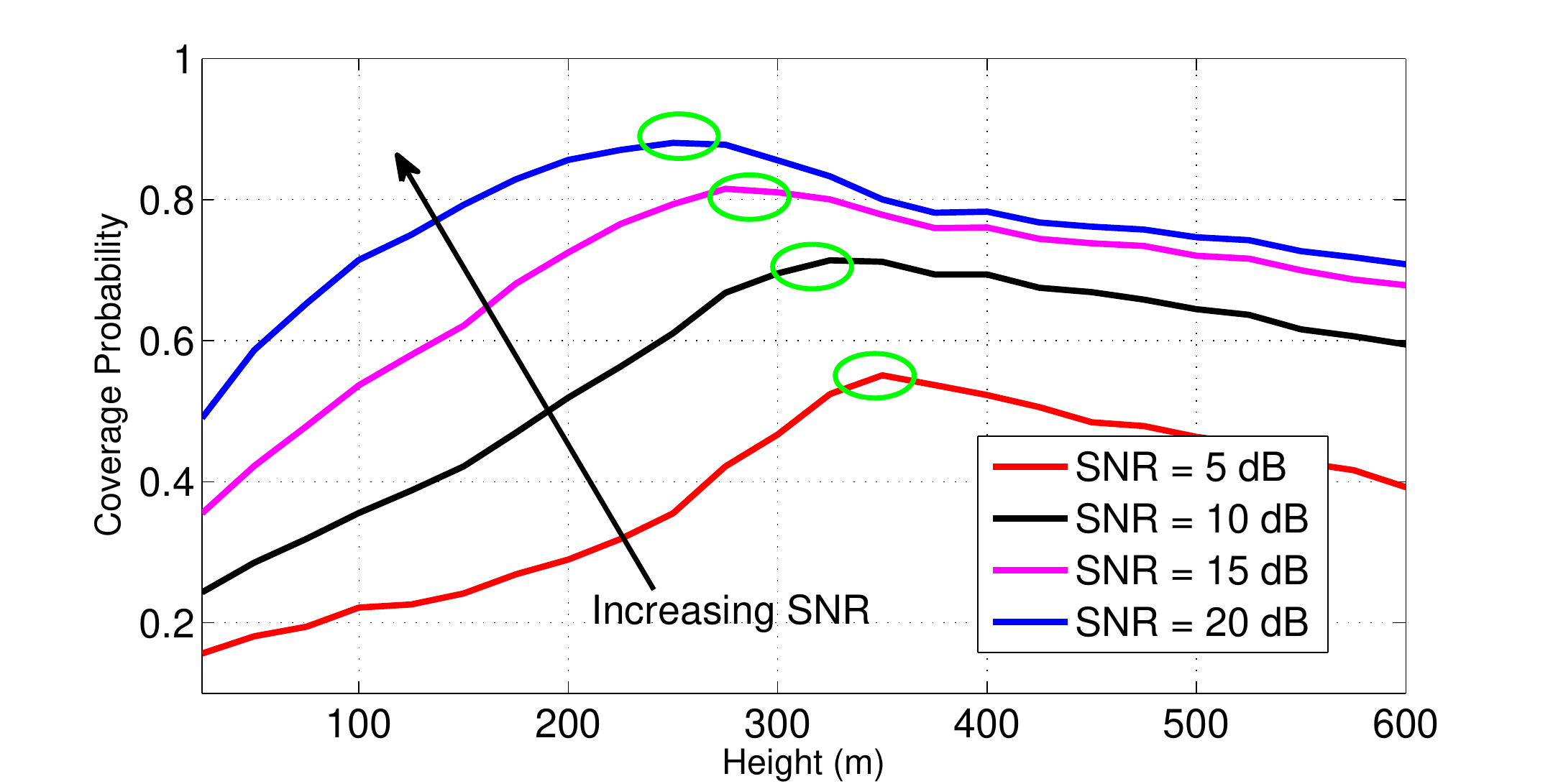}
    \caption{Simulation based on  SUI model in different SNR, condition: $\theta$ = -15 dB, $\lambda$ = 1$/\textrm{km}^2$, $d_0$ = 100 m.}
    \label{fig71}
    \vspace{-3mm}
\end{figure}

Fig. \ref{fig7} shows the simulation result in different SINR thresholds with SNR = 0 dB, and the optimal height is about 350 m where the PLE is almost 2, which implies that the optimal height is fixed and the communication condition is the best with not so much pathloss. Similarly, from the theoretical expression (\ref{eq1})(\ref{eq123}), we also find that the optimal height has nothing to do with the UAV density and the SINR threshold. However, if the SNR is neither too high nor too low, the optimal height will be correlated to other parameters. Fig. \ref{fig71} shows the simulation result in different SNR with $\theta$ = -15 dB, and the optimal height degrades with an increase in the SNR, since the interference begins to affect the coverage probability.

\subsection{Optimal UAV density}

Considering the impact of PLE and altitude, the UAV deployment should be variable in different heights, but the performance has a similar property in distribution density. With the high SNR, the higher density will lead to more interference so that the SINR degrades hence the coverage. However, the serving ABS is closer to the UE when the density increases, thereby improving the SINR hence the coverage in the low SNR condition. Since the signal power and the interference will influence the coverage in different direction, there should be an optimal distribution density with a medium SNR.

\begin{figure}\centering
    \includegraphics[width=3.7in]{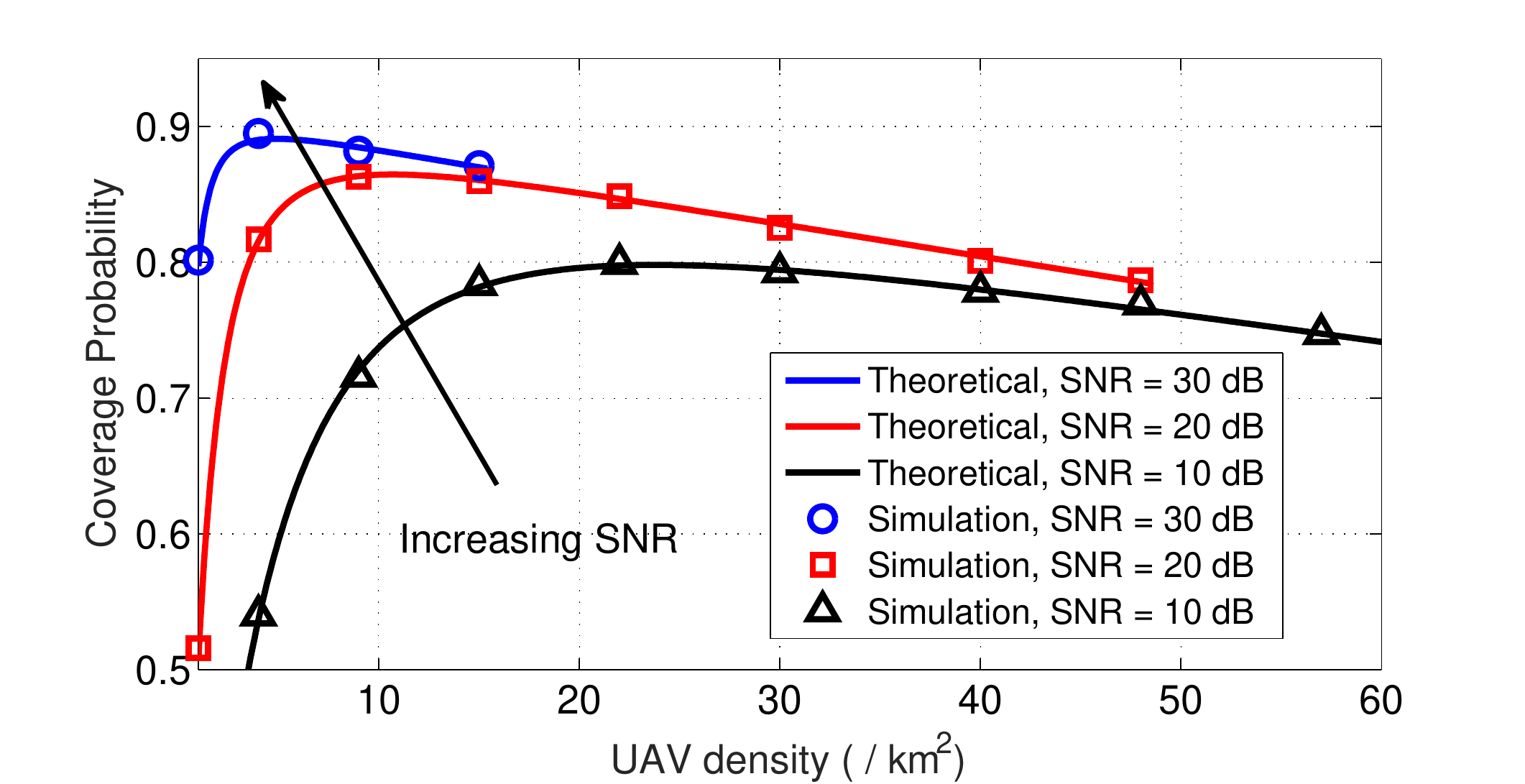}
    \caption{Comparison between different SNR, condition: PLE = 4, height = 100 m, $\theta = - 10$ dB, $d_0$ = 100 m.}
    \label{fig8}
    \vspace{-3mm}
\end{figure}

Fig. \ref{fig8} shows the theoretical and simulation results in different distribution densities with a fixed height. It could be observed that the coverage probability increases as the density increases firstly and then degrades linearly. Because at the beginning the density is very low and its increase could improve the signal power significantly, but if the density reaches a certain value, the interference will begin to have a great impact on the channel and degrades such performance.

The exact expression for the optimal density is also derived below. The series expansion of $Q(x)$ for larger $x$ is \cite{qfunction}
\begin{equation} \label{eq13}
Q(x) = \frac{\textrm{exp}(-\frac{x^2}{2})}{\sqrt{2\pi}\sqrt{1+x^2}},
\vspace{0mm}
\end{equation}
which allows simplification of (\ref{eq12}) to
\begin{equation} \label{eq14}
\begin{aligned}
\textbf{P}(\theta,z)
& \approx \frac{\lambda \pi d_0^2}{\sqrt{2\theta \beta_0}} e^{-\lambda \pi \rho z^2 - \frac{\theta \beta_0 z^4}{d_0^4}}\cdot \frac{1}{\sqrt{1+(\kappa+\frac{z^2}{d_0^2}\sqrt{2\theta \beta_0})^2}} \\
& \approx a \lambda \cdot e^{-c\lambda} \cdot \frac{1}{\sqrt{1+b \lambda^2 }},
\vspace{0mm}
\end{aligned}
\vspace{0mm}
\end{equation}
when setting
\begin{equation} \label{eq15}
\left\{
\begin{aligned}
&a = \frac{d_0^2 }{\sqrt{2\theta \beta_0}} e^{\frac{-\theta \beta_0 z^4}{d_0^4}} \\
&b = \frac{\pi^2(1+\rho)^2 d_0^4}{2\theta \beta_0}, \\
&c = \pi \rho z^2 \\
\end{aligned}
\right.
\vspace{0mm}
\end{equation}
in the case of high SNR and large SINR threshold. When making the differential function of (\ref{eq14}) to be 0, the optimal density to maximize the coverage probability is given as below,
\begin{equation} \label{eq16}
\frac{\partial \textbf{P}}{\partial \lambda} = \frac{ae^{-cx}(-bcx^3-cx+1)}{(bx^2+1)^{3/2}}=0,
\vspace{0mm}
\end{equation}
\begin{equation} \label{eq17}
\lambda_{opt} = \sqrt[^3\!]{\frac{1}{2bc}+\sqrt{\frac{1}{4b^2 c^2}+\frac{1}{8b^3}}}+\sqrt[^3\!]{\frac{1}{2bc}-\sqrt{\frac{1}{4b^2 c^2}+\frac{1}{8b^3}}}.
\vspace{0mm}
\end{equation}
Plugging (\ref{eq15})(\ref{eq17}) into (\ref{eq12})(\ref{eq121}) gives the optimal coverage probability with the approximate method.

\begin{figure}\centering
    \includegraphics[width=3.6in]{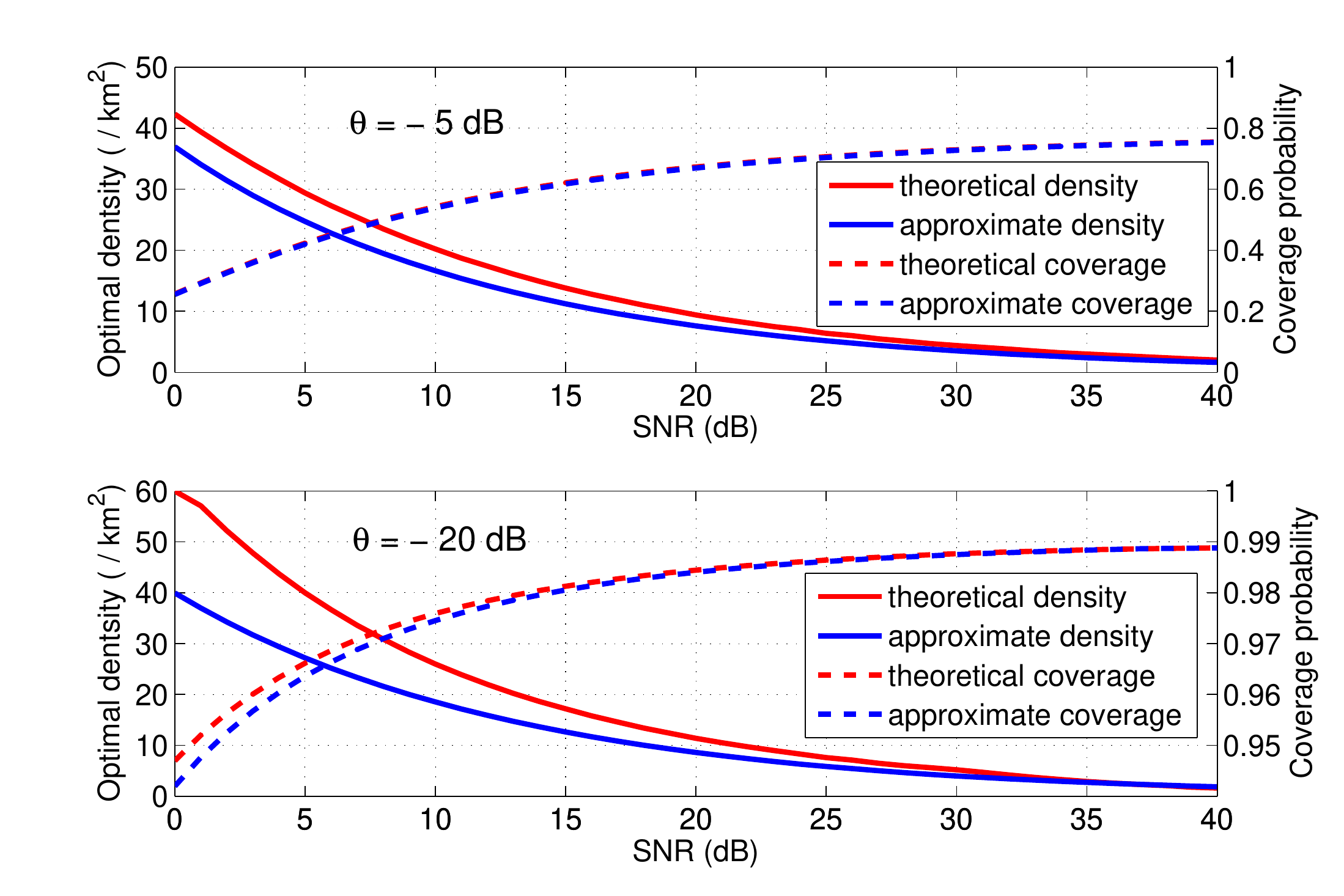}
    \caption{Theoretical and approximate results of optimal density and coverage probability, condition: PLE = 4, height = 100 m , $d_0$ = 100 m.}
    \label{fig9}
    \vspace{-3mm}
\end{figure}

Fig. \ref{fig9} shows the results of the optimal density and coverage probability in two methods, and the theoretical results are obtained from simulation based on (\ref{eq12})(\ref{eq121})(\ref{eq122}). As expected, the approximate density matches the theoretical result well in the case of high SNR and large SINR threshold, since the variable in $Q$ function is large enough. Furthermore, the approximate coverage could be used to approach the theoretical result significantly because of less difference between them.

\section{Conclusion}

This paper provides explicit expressions and detailed analysis for the average coverage probability of the UAV-assisted cellular networks in urban environments. By modeling the UAV as a PPP in a certain height, we derived an exact expression for coverage probability for the Rayleigh fading channel. The property of coverage probability is also analyzed based on the AG channel, and we reveal some important trends in terms of the UAV height, PLE and UAV density, in the case of high SNR and low SNR, respectively. With the low SNR, there is an optimal height to maximize the coverage probability based on the SUI model. The exact expression of the optimal density in a certain height is also derived, and the result in different height could be computed in a similar method. The work in this paper could be used to design the UAV deployment to implement a good coverage in the UAV-assisted cellular networks.

\section*{Acknowledgement}

The research presented in this paper has been kindly funded by the projects as follows, National S$\&$T Major Project (2017ZX03001011), China's 863 Project (2015AA016202), National Natural Science Foundation of China (61631013), Foundation for Innovative Research Groups of the National Natural Science Foundation of China (61621091), Tsinghua-Qualcomm Joint Project, Future Mobile Communication Network Infrastructure Virtualization and Cloud Platform (2016ZH02-3), Tsinghua Fudaoyuan Research Fund.

\end{document}